\documentclass[%
 reprint,
superscriptaddress,
 amsmath,amssymb,
 aps,
prstab,
nofootinbib
]{revtex4-1}

\usepackage{scrextend}
\usepackage{graphicx}
\usepackage{dcolumn}
\usepackage{bm}
\usepackage{amsmath}
\usepackage[colorlinks=true]{hyperref}

\begin{document}
\begin{abstract}
A proposed experiment to demonstrate Optical Stochastic Cooling (OSC) in the Cornell Electron Storage Ring (CESR) based on an arc-bypass design is presented. This arc-bypass provides significantly longer optical delay than the dog-leg style chicane, opening up the possibility of a multi-pass or staged optical amplifier that can achieve the gains required for effective cooling of hadron or heavy-ions. Beyond introducing the arc-bypass, in this paper we study the stability requirements for the dipoles comprising it and investigate the use of an optical feedback system to relax the dipole and light-path stability tolerances.
\end{abstract}
\title{Optical Stochastic Cooling with an Arc-Bypass in CESR}
\author{M.B. Andorf, W.F. Bergan, I.V. Bazarov, J.M. Maxson, V. Khachatryan, D.L. Rubin, S.T. Wang}
\affiliation{CLASSE, Cornell University, 161 Synchrotron Drive, Ithaca, New York 14853-8001, USA} 
\maketitle

\section{Introduction}
Optical Stochastic Cooling (OSC) is a proposed particle beam cooling technique that extends the widely implemented stochastic cooling\cite{Mohl} from microwave to optical frequencies~\cite{OSC_Mikhailichenko}. Typically, the transition to optical frequencies is via a pair of undulators; the ``pickup undulator" (PU) and the ``kicker undulator" (KU) and is motivated by an approximate 4-orders of magnitude increase in cooling bandwidth. In the transit-time method of OSC~\cite{OSC_Zolotorev, SYLee}, a particle emits a wave-packet in the upstream PU, and interacts with that same (amplified) wave-packet in the downstream KU. The result of the interaction is an energy exchange (kick) to the particle, the sign and magnitude of which depend on the relative transit-time of the particle (determined by the magnetic particle bypass) and the wave-packet (determined by the optical light path). The lengths of the two paths are adjusted so that the reference particle receives no kick while the arrival of a generic test particle is advanced or delayed in such a way that the kick reduces its Courant-Snyder invariant and/or momentum offset. The momentum kick can damp longitudinal motion directly, and with a suitable introduction of dispersion in the PU and KU, horizontal as well. In any stochastic cooling scheme the maximum achievable damping rate for a collection of particles is determined through a trade off between the amplitude of the corrective kick through the process just described, and the incoherent kicks that it receives from its neighbors within a longitudinal slice of the beam, of width $\Delta t\approx1/2\Delta f$ where $\Delta f$ is the frequency bandwidth of the system.

State-of-the-art microwave-based stochastic cooling systems are limited to about 8 GHz bandwidth~\cite{Pasquinelli}. For the dense particle beams found in hadron and heavy-ion colliders at collision energy, this limits the damping rate to significantly less than the growth rates from beam heating effects like Intra-beam Scattering (IBS), and is therefore ineffective during a beam store. Consequently, there has been much interest in developing alternative cooling techniques including Coherent electron Cooling (CeC)\cite{CEC1}, Microbunched Electron Cooling (MBEC)\cite{MBEC1}, and of course, OSC---all of which rely on the transit-time method and aim to increase the cooling bandwidth.
An estimate of the bandwidth of undulator radiation is $\Delta f \approx c/(N_u \lambda_l)$, where $N_u$ is the number of periods, $\lambda_l$ is the zero-angle wavelength, and can therefore exceed 10's of THz for resonant optical wavelengths. Thus with OSC, the width of the longitudinal slice and, correspondingly, the number of incoherent kicks a particle receives is greatly reduced and in principle can increase the damping rate, compared to ordinary stochastic cooling, by several orders of magnitude. 

Currently there are two complementary programs for developing OSC, one at Fermilab in the Integrable Optics Test Accelerator (IOTA) \cite{Jarvis, IOTAVal} with 100 MeV electrons and the other in the Cornell Electron Storage Ring (CESR) at Cornell University with 1 GeV electrons. 

In the IOTA demonstration a dog-leg style chicane consisting of 4 dipoles and a center defocusing quadrupole occupies one straight section of the ring. In this configuration the particle path is horizontally displaced to make way for the light-path, which follows a straight line from the PU to the coaxial KU. Refractive optics are used to both focus the PU light into the KU and delay it to compensate for the additional distance traveled by the particle beam through the chicane, $\Delta L$. 

In the dog-leg bypass, $\Delta L$ scales inversely with the cooling ranges~\cite{OSC_val} which typically constrains $\Delta L$ to be limited to a few millimeters(see next section). For electrons at 100 MeV, even with passive cooling (where no Optical Amplifier (OA) is present), the OSC damping rates can be made to greatly exceed damping due to synchrotron radiation. Thus, the IOTA experiment will be an elegant and clear proof-of-principle demonstration of OSC's working principles applied to electrons. However, for cooling of high-energy hadrons a passive scheme is too slow, where it is estimated that 20-30 dB of gain will be required from an OA to effectively combat emittance growth from IBS~\cite{OSC_val}. For IOTA's active test of the OSC, a 2-mm thick Chromium Zinc Selenide (Cr:ZnSe) crystal with an amplification peak at 2.45 $\mu$m is being considered\cite{Amplifier_Andorf}. The amplification of this crystal is limited due to a depletion of the ground-state ions and associated thermal effects from absorption of the pumping laser power. Consequently, the amplifier yields a predicted gain of only 7~dB.

An OA based on a Titanium-Sapphire crystal (Ti:Sapph) has also frequently been considered for OSC~\cite{Amplifier_Zholents,IOTAVal}. For the dog-leg chicane, accounting Ti:Sapph's shorter central wavelength of 790 nm, the total optical delay must be reduced proportionately in order to maintain the same cooling ranges. This  results in a crystal length of less than a millimeter making high gain amplification difficult to achieve.
\begin{figure}
	\centering
	\includegraphics*[width=0.45\textwidth]{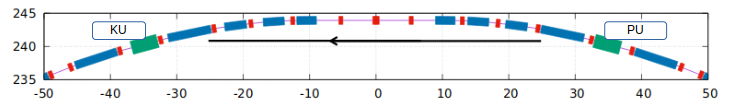}
	\caption{Schematic of the arc-bypass in CESR. Light is extracted from the vacuum chamber in the immediate downstream dipole (blue) of the PU (green) and reenters in the immediate upstream dipole of the KU. Dimensions are in m.} 
\label{Fig:arc}
\end{figure}
For both gain mediums, the very limited optical delay characteristic of the dog-leg style chicane limits the achievable amplification.

Thus by contrast, the implementation of OSC in CESR is based on an alternative magnetic bypass where the relative delay of light and particle beam is \textit{independent} of the cooling ranges; we refer to our design as an arc-bypass. The light path is along a chord that intercepts $30^\circ$ of the ring arc. The vacuum chamber in the dipole just downstream of the PU is outfitted with an in-vacuum pick-off mirror to extract the PU light. A second in-vacuum mirror, in the dipole chamber just upstream of the KU, directs the light along the path of the particle beam in the KU. See Fig.~\ref{Fig:arc}. The distance from midpoint of PU to KU is 71.78 m and $\Delta L$ is approximately 20 cm; thus opening up the possibility of multi-pass or staged amplification schemes, so that gains needed for hadron or heavy-ion cooling are achievable with conventional laser amplification techniques. 

The paper is organized as follows: in Section~\ref{Sec:BasicFormulas} we briefly review the relevant theory before presenting the lattice design of the arc-bypass in Section~\ref{Sec:ArcDesign}. We then analyze the stability requirement of dipole fields in the ring and consider use of a feedback system for stabilization in Sections~\ref{Sec:Stability} and ~\ref{Sec:Feedback}. Finally, we present multi-particle tracking simulations of the cooling process in Section~\ref{Sec:Sims}.
\section{Overview of OSC Particle Dynamics}
\label{Sec:BasicFormulas}
A general theory of OSC working principles can be found in reference \cite{OSC_val}. Here we reproduce some of the major formulas relevant to our discussion.

A particle arriving in the PU with coordinates $x$, $x'$ and $\Delta P/P$, will have a path-length difference, with respect to the reference particle, while traveling between the PU and KU
\begin{equation}
\Delta s =M_{51}x+M_{52}x'+M_{56}\Delta P/P.
\end{equation}
It will receive an energy kick
\begin{equation}
\delta u=\Delta \mathcal{E}\sin(k_l\Delta s)
\label{deltau}
\end{equation}   
where $M_{5,n}$ are elements of the 6x6 transfer matrix from PU to KU centers, $\Delta \mathcal{E}$ is the kick amplitude which is determined by the undulator parameters, particle beam energy, light optics design and OA gain \cite{Andorf_NIM,Andorf_PRAB}, while $k_l\equiv 2\pi/\lambda_l$ where $\lambda_l$ is the zero-angle wavelength of the undulator radiation.

The path length from PU to KU through the OSC bypass will depend on the betatron and synchrotron phases and amplitudes of the particle at the PU, as well as the matrix elements $M_{5,n}$. Throughout the cooling process a particle will oscillate longitudinally with respect to the arrival time of its PU wave-packet in the KU. The longitudinal amplitude $s_x$ of this motion arising from the particle's Courant-Snyder invariant $\epsilon$\footnote{The Courant-Snyder invariant is defined as $\epsilon=\gamma x^2+2\alpha x x'+\beta x'{^2}$.} is
\begin{equation}
s_x=\sqrt{\epsilon\big(\beta_{PU} M_{51}^2-2\alpha_{PU} M_{51}M_{52}+\gamma_{PU} M_{52}^2\big)}
\label{sx}
\end{equation}
where $\beta_{PU}$, $\alpha_{PU}$ and $\gamma_{PU}$ are the lattice Twiss parameters evaluated at the PU.
The longitudinal amplitude $s_p$ due to the particle's synchrotron amplitude, $\big(\frac{\Delta p}{p}\big)_m$ is:
\begin{equation}
s_p=(M_{51}D_{PU}+M_{52}D_{PU}'+M_{56})\big(\frac{\Delta P}{P}\big)_m,
\label{su}
\end{equation}
where $D_{PU}$ is the lattice dispersion and $D_{PU}'$ is its derivative at the PU. For small particle amplitudes satisfying $k_ls_x, k_ls_p\ll1$, the horizontal and longitudinal damping rates are
\begin{equation}
\lambda_{xo}=-k_l\frac{M_{51}D_{PU}+M_{52}D_{PU}'}{2\tau_s}\frac{\Delta \mathcal{E}}{U_s}
\label{DampingX}
\end{equation}
and
\begin{equation}
\lambda_{po}=k_l\frac{M_{51}D_{PU}+M_{52}D_{PU}'+M_{56}}{2\tau_s}\frac{\Delta \mathcal{E}}{U_s}
\label{DampingU}
\end{equation}
where $\tau_s$ and $U_s$ are the revolution period and energy of the beam, respectively. For large amplitude particles the sinusoidal nature of the kick must be taken into account resulting in amplitude dependent damping rates
\begin{equation}
\lambda_x=2\lambda_{xo}\frac{J_0(k_ls_p)J_1(k_ls_x)}{k_ls_x}
\label{DampingLargeX}
\end{equation}
and
\begin{equation}
\lambda_p=2\lambda_{po}\frac{J_0(k_ls_x)J_1(k_ls_p)}{k_ls_p}.
\label{DampingLargeU}
\end{equation}
In the above we see that 1-dimensional damping (i.e. $k_ls_p\ll 1$ for horizontal damping) requires a particle longitudinal amplitude, expressed in units of undulator radiation phase, to be less than $\mu_{1,1}\approx 3.83$. Simultaneous damping in both planes requires both $k_ls_x$ and $k_ls_p < \mu_{0,1}\approx 2.41$, where $\mu_{n,m}$ is the $m^{th}$ zero of the $J_n$ Bessel function. We define the emittance acceptance 
\begin{equation}
\epsilon_{max}=\frac{\mu_{0,1}^2}{k_l^2(\beta_{PU} M_{51}^2-2\alpha_{PU} M_{51}M_{52}+\gamma_{PU M_{52}})}
\end{equation}
and momentum acceptance as
\begin{equation}
\big(\frac{\Delta p}{p}\big)_{max}=\frac{\mu_{0,1}}{k(M_{51}D_{PU}+M_{52}D_{PU}'+M_{56})}
\end{equation}
Because it is possible for a particle to be momentarily heated in one plane, while damped in the other, before eventually damping in both planes (see \cite{Andorf:Thesis} for complete details), the $\mu_{0,1}$ boundary gives a conservative estimate of the acceptances. Finally, the above expressions can be used to obtain the cooling ranges
\begin{equation}
\eta_x=\sqrt{\frac{\epsilon_{max}}{\epsilon_o}}\quad \quad \eta_p=\frac{1}{\sigma_p}\big(\frac{\Delta P}{P}\big)_{max}.
\label{CoolingRanges}
\end{equation}
where $\epsilon_o$ and $\sigma_p$ are, respectively, the horizontal beam emittance and longitudinal momentum spread.

In reference~\cite{OSC_val} the above formulas are applied to the dog-leg style chicane. Because the cooling ranges are inversely proportional to the total optical delay, for beams with nanometer scale emittance and order 0.01\% energy spread, the optical delay can be no more than a few millimeters in order to have sufficiently large cooling ranges.
\section{Arc-bypass design}
\label{Sec:ArcDesign}
\begin{figure}
	\centering
	\includegraphics*[width=0.48\textwidth]{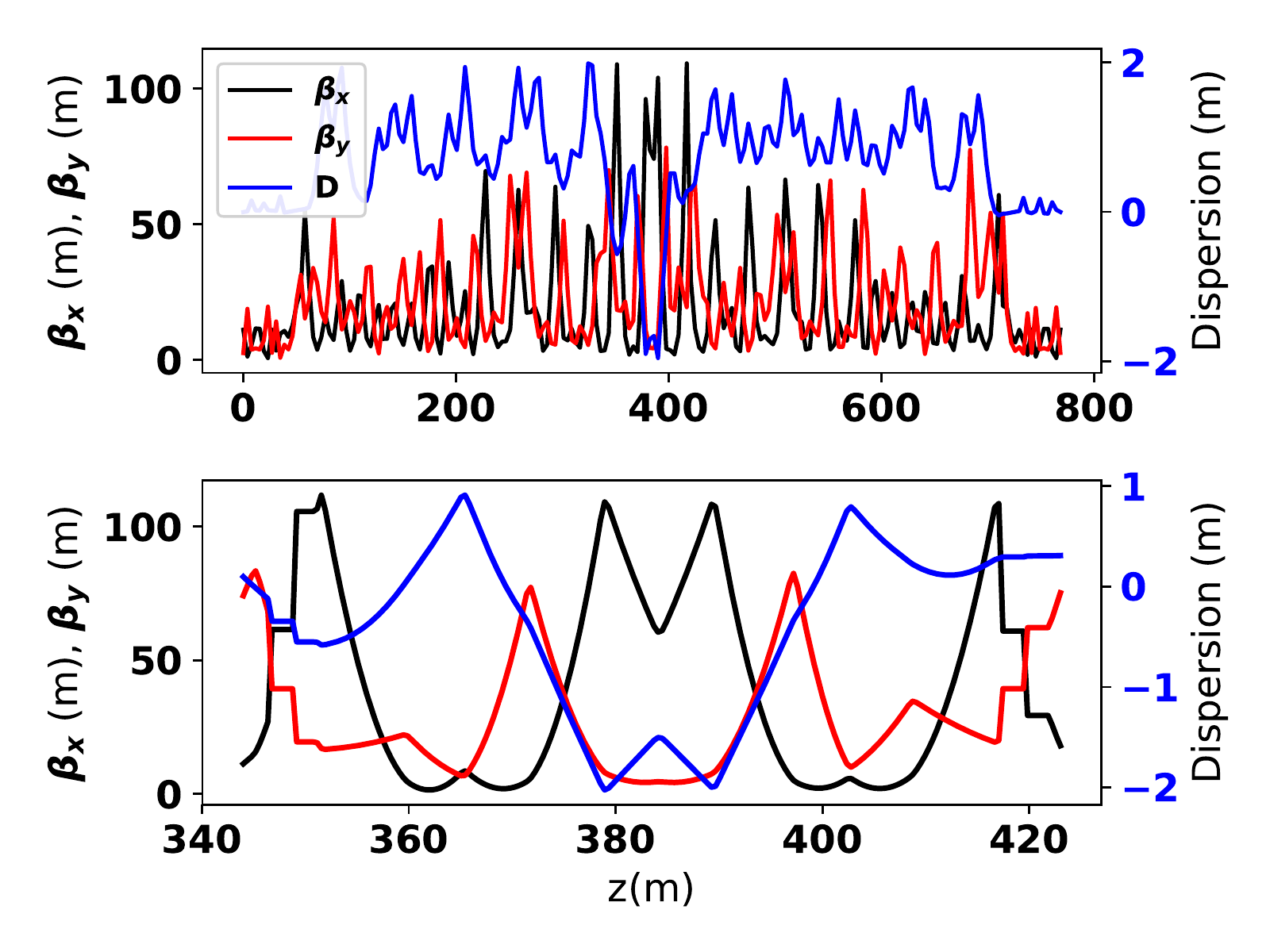}
	\caption{The lattice functions throughout CESR (top) and zoomed in on the arc-bypass (bottom). The PU and KU centers are located at 348 m and 420 m, respectively. }. 
\label{Fig:Lattice}
\end{figure}
In order to obtain the linear optics needed for OSC, we used the optimization routines made available in the Tool for Accelerator Optics (\texttt{TAO})\cite{Tao} program, and the more general formulas of the previous section; more details can be found in \cite{Bergan_IPAC}.

In general, there is a trade-off between the size of the cooling ranges and the total damping rate, $\lambda_{xo}+\lambda_{po}$, and a further trade off between the horizontal and longitudinal damping rates. In view of the relatively fast synchrotron radiation damping rates, $\lambda_{x,SR}$ and $\lambda_{p, SR}$, at 1-GeV in CESR, for the CESR demonstration we choose to optimize horizontal damping. However, we have also designed bypass optics that provide simultaneous longitudinal and horizontal cooling, thus demonstrating the flexibility of our configuration. The parameters of the arc-bypasses are given in Table.~\ref{tab:CoolParams} and the lattice functions for the lattice optimized for horizontal cooling are shown in Fig.~\ref{Fig:Lattice}.

The PU and KU are helical undulators comprised of 14 periods of length 28 cm, yielding an undulator parameter $K=4.51$ and a zero-angle wavelength of 780 nm. Using the formula for the theoretical kick amplitude $\Delta \mathcal{E}$ found in \cite{Andorf_PRAB} we obtain a value of 420 meV in the absence of amplification. For this value we have assumed a lossless light transport system consisting of a telescope that provides 1-to-1 imaging along the lengths of the PU and KU, and a normalized angular acceptance $\gamma\theta_m=3.5$, where $\theta_m$ is the angle subtended by the telescope. With this kick, the horizontal damping rate is marginally faster than that from synchrotron radiation damping at 1 GeV\footnote{The combined OSC and synchrotron damping rates in this case is substantially faster than the synchrotron damping rate and therefore results in measurable cooling of the beam.}.

As in the dog-leg chicane \cite{VAL_COOL15, Kafka:Thesis}, nonlinear path lengthening is corrected with sextupoles within the arc bypass. The distribution of sextupoles outside of the bypass is optimized for dynamic aperture and to compensate chromaticity.

\begin{table}[h!]
  \begin{center}
  \caption{Major cooling parameters for the arc-bypass in CESR.}
  \label{tab:CoolParams}
    \begin{tabular}{l|c|c} 
      \textbf{Parameter} & \textbf{Horizontal Cooling} &\textbf{Simultaneous Cooling} \\
      \hline
      $\epsilon_o (nm)$ & 0.73 & 0.73 \\
      $\sigma_p$ &3.7$\times 10^{-4}$ & 3.7$\times 10^{-4}$\\
      $\eta_x$ & 2.8 & 3.11\\
      $\eta_p$ & 31.3 & 2.1 \\
      $\lambda_{xo}$ (s$^{-1}$) & 0.91 & 0.77 \\
      $\lambda_{x,SR}$ (s$^{-1}$)& 0.73 & 0.73\\
      $\lambda_{po}$ (s$^{-1}$)& 0.01 & 0.24\\
      $\lambda_{p, SR}$ (s$^{-1}$)& 1.27 & 1.27\\
      \hline
    \end{tabular}
  \end{center}
\end{table}
\section{Path-length Stability Requirements for OSC}
\label{Sec:Stability}
\begin{figure}
	\centering
	\includegraphics*[width=0.49\textwidth]{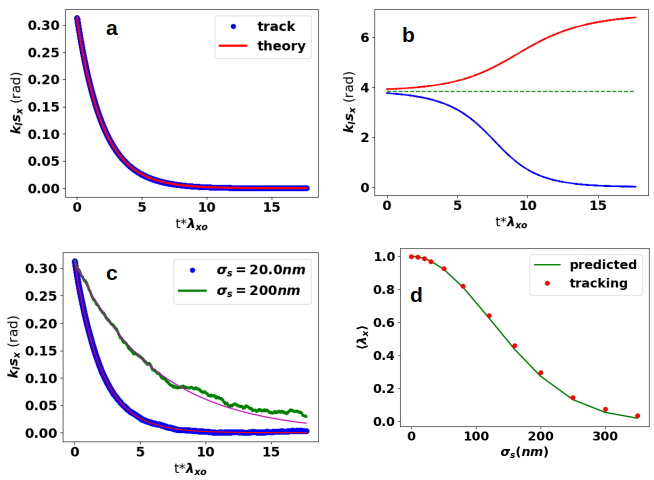}
	\caption{Results from the fast tracking routine: \textbf{(a)} demonstrates agreement between theoretical damping rate and tracking for a small amplitude particle while \textbf{(b)} verifies a reduction in the damping rate at large amplitude and a cooling boundary at $J_1(\mu_{1,1})$. \textbf{(c)} shows examples of the reduction in the damping rate when a path-length jitter is applied as well as exponential best fits used to extract the reduced rates. Finally, \textbf{(d)} plots the reduced damping rate for a given RMS path-length jitter from tracking with the prediction from Eq.~\ref{RedDamping} }. 
	\label{Fig:FastTracking}
\end{figure}
OSC requires extreme (sub-optical wavelength) accuracy in the relative particle and light path-lengths \cite{ZholentsErrors}. For the arc-bypass,  as compared to the dog-leg chicane, the path-lengths from PU to KU for both charged particles and light are much longer and there is significantly more bending in the bypass. These considerations motivated an analysis of the required stability for the guide field dipoles both within and outside of the bypass. In particular, dipole field errors cause a change in the path length of the reference orbit and, therefore, a path-length error, $\Delta s_{err}$. We consider two cases: (i) a coherent change to all dipole field strengths in the ring by the same relative amount at the same instant, and (ii) random incoherent changes in the field strength of each dipole separately.

The effect of $\Delta s_{err}$ on cooling depends on the time scale over which the errors occur. For changes on a time scale much longer than the damping time, the rates in Eq's.~\ref{DampingX} and \ref{DampingU} are reduced by a constant factor $\cos(k_l\Delta s_{err})$. For errors that occur on a time scale much shorter than the damping time, an average over the instantaneous reduction in the damping rate, $\lambda_{xo,po}\cos(k_l\Delta s_{err})$, is needed. For example, if the path length error is a Gaussian-random variable with an RMS-spread $\sigma_s$, the reduced damping rate is
\begin{align}
\langle \lambda_{x,p}\rangle=& \nonumber \\
&\lambda_{xo,po}\int_{-\infty}^{\infty}\cos(k_l\Delta s_{err})\frac{\exp(-\Delta s_{err}^2/2\sigma_s^2)}{\sigma_s\sqrt{2\pi}}d\Delta s_{err},
\end{align}
which simplifies to
\begin{equation}
\label{RedDamping}
\langle \lambda_{x,p}\rangle=\lambda_{xo,po}\exp(-k_l^2\sigma_s^2/2).
\end{equation}
To confirm Eq.~\ref{RedDamping}, a fast tracking method was implemented using transverse transfer matrices obtained in \texttt{TAO} to simulate horizontal, single-particle damping as follows: A particle is placed at the PU with some initial betatron coordinates $(x_{\beta,PU},x_{\beta,PU}')$ and propagated to the KU. At the KU an energy kick is applied resulting in a change to the particle's betatron coordinates\footnote{Note there is no change to the particle's geometric coordinates during the kick with the two related as $x=x_\beta +D\delta u/U$}:
\begin{equation}
\Delta x_\beta=-\frac{\Delta \mathcal{E}}{U_s}D\sin(k_l(M_{51}x_{\beta,PU}+M_{52}x_{\beta,PU}'))
\end{equation}
and
\begin{equation}
\Delta x'_\beta=-\frac{\Delta \mathcal{E}}{U_s}D'\sin(k_l(M_{51}x_{\beta, PU}+M_{52}x_{\beta, PU}')).
\end{equation}
The updated coordinates at the KU are then propagated around the ring to the PU and the entire process is repeated.

The above scheme neglects changes to the particle's longitudinal coordinates imparted by the energy kick which remain at $(0,0)$ throughout the tracking. This is a reasonable approximation since over the course of a single betatron oscillation the net energy kick nearly averages to zero and the total accumulated energy change of the particle remains small throughout the damping process. 

With this method we first confirmed a particle with a small longitudinal amplitude damps as expected according to Eq.~\ref{DampingX}. In Fig.~\ref{Fig:FastTracking}\textbf{a} the amplitude of the particle's displacement $s_x$ is computed using Eq.\ref{sx} at each turn and compared to the function $s_x(t)=s_{xo}\exp(-\lambda_{xo}t)$, which is seen to be in good agreement with the tracking. Next in Fig.~\ref{Fig:FastTracking}\textbf{b} we confirm the alteration to the damping predicted in Eq.~\ref{DampingLargeX}. Namely, (i) that a particle with a large amplitude that initially satisfies $k_ls_x<\mu_{1,1}$ will still damp, albeit at a slower rate, as shown with the blue curve and (ii) a particle with an amplitude exceeding $\mu_{1,1}$ will anti-damp to $\mu_{1,2}\approx 7.01$ shown with the red curve. 

We then applied turn-by-turn Gaussian-random path-length jitter to a small amplitude particle. The damping rate is evidently reduced and an exponential best fit is performed to obtain the reduced rate as shown in Fig.~\ref{Fig:FastTracking}\textbf{c}. Finally, the reduction is found to be in good agreement with the Eq.~\ref{RedDamping} as shown in Fig.~\ref{Fig:FastTracking}\textbf{d}.

\subsection{Coherent Dipole Errors}
A coherent error on all dipoles is equivalent to a change in the beam energy. Therefore, distortions of the reference orbit will be proportional to the local dispersion function,
\begin{equation}
 \big(x(z),x(z)'\big)=R_{err}\big(D(z),D'(z)\big) 
 \end{equation}
where $R_{err}\equiv \Delta \rho/\rho$ and $\rho$ is the bending radius of the dipole. There are two contributions to the change in the path length of the reference particle: (i) a transverse displacement of the equilibrium orbit at the PU, so that $\Delta s_1=R_{err}(M_{51}D_{PU}+M_{52}D'_{PU})$ and (ii) a direct change in the path length from PU to KU:
\begin{equation}
\Delta s_2=\int_{PU}^{KU}\frac{R_{err}D(z)}{\rho}dz=R_{err}M_{56}.
\end{equation}
Thus the total path-length change is
\begin{equation}
\Delta s_{dip}=R_{err}(M_{51}D_{PU}+M_{52}D'_{PU}+M_{56}).
\label{DeltaSCoh}
\end{equation}
The above expression is identical to Eq.~\ref{su} with $\big(\Delta p/p\big)_m$ replaced by $R_{err}$. For a coasting beam this correspondence implies that dipole sensitivity requirements scale proportionately with the longitudinal cooling range. 

So far we have neglected any interaction with the RF system. With a coherent decrease in dipole strengths around the ring, the reference particle, in order to stay synchronous with the RF cavity, decreases in energy by the same relative amount and, consequently, there is an additional change to the path-length, $\Delta s_{rf}=-R_{err}(M_{51}D_{PU}+M_{52}D'_{PU}+M_{56})$ exactly cancelling the path change from the dipoles, $\Delta s_{coh}=\Delta s_{dip}+\Delta s_{rf}=0$. Thus, the reference particle's transit-time is not affected. The beam centroid, however, no longer coincides with the reference particle's energy and the beam will oscillate  around the new reference energy. The damping of this oscillation is similar to single particle damping and thus happens at a rate (from OSC) $\lambda_{p,centroid}=2\lambda_{po}J_1(k_l\Delta s_{dip})/k_l\Delta s_{dip}$. Practically speaking this effect is negligible in the CESR test because of the large momentum acceptance. For example, coherent relative dipole change as large as $10^{-3}$ would result in a displacement less than $30$ nm.

There is an additional nonlinear path lengthening of the reference particle through the bypass
\begin{equation}
\begin{aligned}
\Delta s_{NL}=&\int_{PU}^{KU}\bigg(1-\sqrt{1+\bigg(\frac{dx}{dz}\bigg)^2}\bigg)dz\\
\quad &\approx \frac{R_{err}^2}{2}\int_{PU}^{KU}D'^2(z)dz.
\label{DeltaSCohNL}
\end{aligned}
\end{equation}
Numerical integration of the above equation yields $\Delta s_{NL}\approx0.68R_{err}^2$ for our bypass, which implies a 1 nm path change for $R_{err}$ of 3.8$\times 10^{-5}$. In the next section we will see that contributions to path-length error from coherent changes are small in comparison to incoherent changes in dipoles and we therefore ignore their effect. 
\subsection{Incoherent Dipole Errors}
\label{Sec:StabilityIncoherent}
It is well known from dipole perturbation theory \cite{Wiedemann}, that a single dipole field error will change the closed orbit such that at the PU
\begin{equation}
x_{err,k}=R_{err,k}\theta_k\sqrt{\beta_{PU}\beta_k}\frac{\cos(\nu\pi-\nu\Delta \phi_k)}{2\sin(\pi\nu)}
\end{equation}
where the subscript $k$ denotes the $k^{th}$ dipole in the ring, $\theta_k$ is its bending angle, and $\Delta \phi_k$ is the phase advance from the dipole to the PU. The orbit distortion results in a path-length error 
\begin{equation}
\label{s0_coh}
\Delta s_{0,k}=M_{51}x_{err,k}+M_{52}x'_{err,k}.
\end{equation}
If the dipole is within the bypass, there will be two additional, direct changes to the particle's path-length. First, as it travels a distance $z_{dip}$ inside the dipole, it is displaced horizontally an amount $\Delta x=R_{err}\theta(z_{dip}) z_{dip}/2$. Integrating over the dipole's length, $L$, yields a path-length change, 
\begin{equation}
\label{s1_incoh}
\Delta s_{1,k}=\int \frac{\Delta x}{\rho} dz_{dip}=R_{err,k}\frac{L_k\theta_k^2}{6}.
\end{equation} 
Then upon exiting the dipole, since the particle has been displaced an amount $(\Delta x,\Delta x')=R_{err}(\theta L/2,\theta)$, there is an additional path change,
\begin{equation} 
\label{s2_incoh}
\Delta s_{2,k}=R_{err,k}(M_{51,dip}\theta_k L_k/2+M_{52,dip}\theta_k), 
\end{equation}
where the transfer elements are from the exit of the dipole to the center of the KU.

\begin{figure}
	\centering
	\includegraphics*[width=0.485\textwidth]{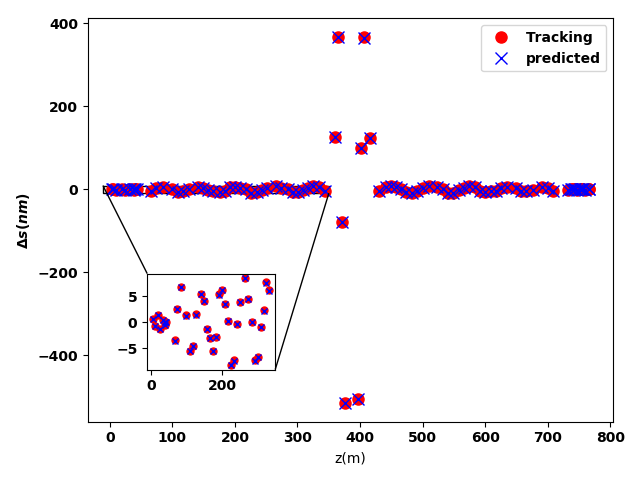}
	\caption{Mean longitudinal displacement computed from particle tracking in \texttt{TAO} for individual dipole errors (red dots) and predicted values (blue x's) assuming a relative error $R_{err}=10^{-5}$. The 8 largest displacements correspond to dipoles inside the arc-bypass.}
	\label{IndividualDeltaS}
\end{figure}
To confirm that the reference particle's longitudinal displacement is given by the summation of Eq's.~\ref{s0_coh}, \ref{s1_incoh} and ~\ref{s2_incoh}, particle tracking in \texttt{TAO} was performed with a single particle placed initially on the reference orbit. A single dipole in the ring is given a relative error $R_{err}=10^{-5}$ and the particle is tracked over 50 turns. This tracking routine is separately repeated for each dipole in the ring. For each dipole the mean longitudinal displacement relative to the unperturbed reference orbit from PU to KU was computed as shown in Fig.~\ref{IndividualDeltaS} and the displacement was found to be in excellent agreement with the above formulas.

We now consider simultaneous fluctuating dipole errors, treating each error as Gaussian-random and uncorrelated, we find that the RMS path-length jitter from dipoles outside the chicane is 40 nm for $R_{err}=10^{-5}$. Assuming the dipoles fluctuate at time scales much faster than the damping rate so that Eq.~\ref{RedDamping} is valid, the damping rate is found to reduce by less than 5~$\%$. By contrast, this same relative error for dipoles inside the bypass results in an RMS path-jitter of 910 nm. Consequently, the damping rate would be essentially reduced to zero. In order to maintain the damping rate to $ 95\%$ of the ideal,  we require a stability of $5\times 10^{-7}$ from dipole fields inside the bypass.

\begin{figure}
	\centering
	\includegraphics*[width=0.485\textwidth]{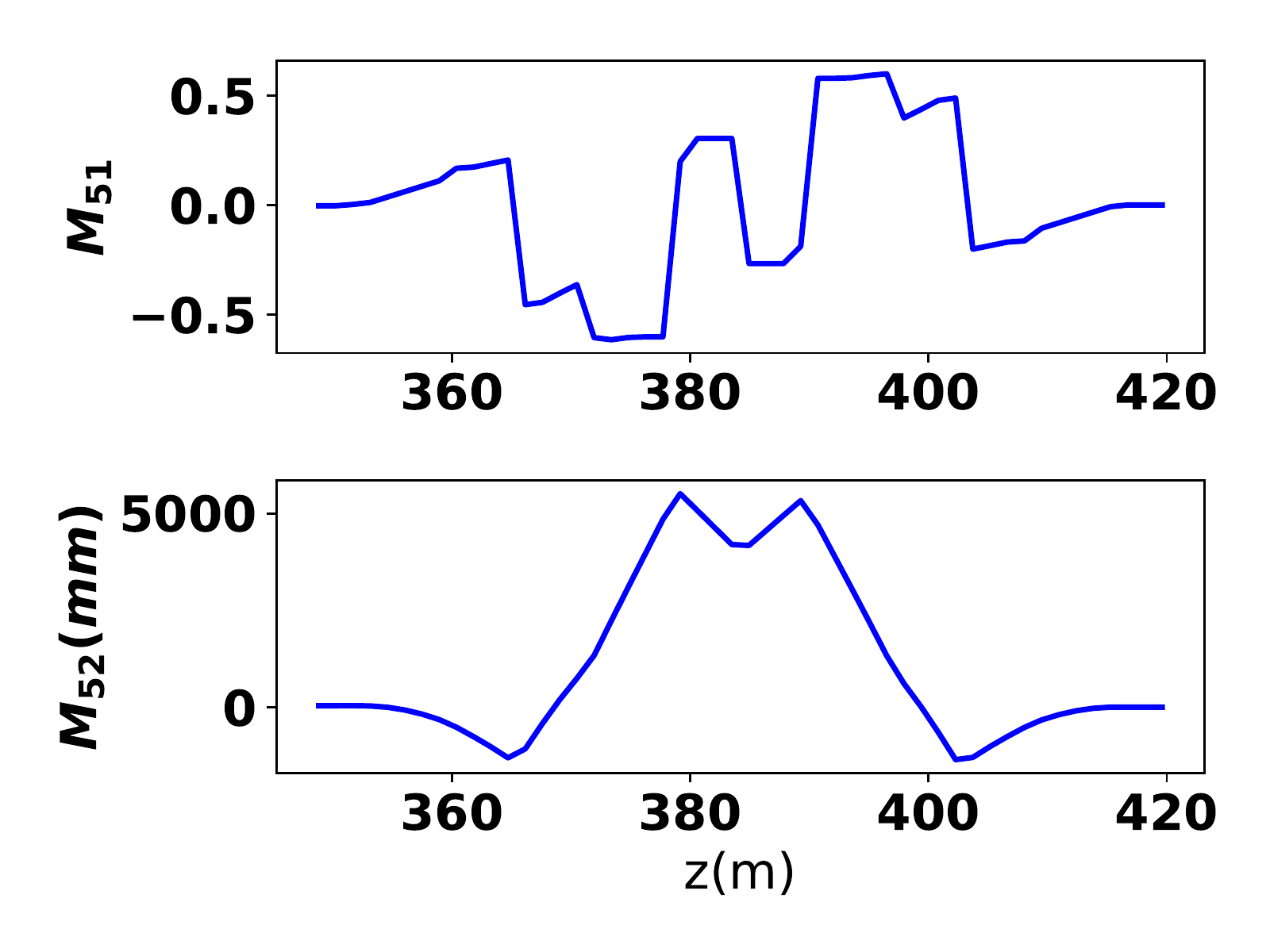}
	\caption{Transfer elements $M_{51}(z)$ and $M_{52}(z)$ computed from a location in the arc-bypass to the KU center.}
	\label{TransferElementFig}
\end{figure}

The sensitivity for bypass dipoles arises partly because $M_{51,dip}$ and $M_{52,dip}$ grow quite large in the bypass as shown in Fig.~\ref{TransferElementFig}, and additionally, because the arc-bypass is comprised of ring dipoles, they have fairly large bending angles and lengths. 
 
Dipole fluctuations, in addition to causing path-length jitter, produce coherent horizontal oscillations of the beam. Because of chromaticity and nonlinearities in the lattice, the oscillations decohere\cite{Meller} and the motion can potentially be translated into an emittance growth. For decoherence arising from a non-linear amplitude dependence on the betatron tune, we find the characteristic time~\cite{Ng} of this process to be 2.2 $s$ in our lattice. This is not significantly longer than the horizontal synchrotron damping rate and we therefore expect a single dipole kick to produce some emittance growth. Given that the emittance in an electron storage ring is reliably predicted by synchrotron radiation alone, and does not need to account emittance growth from dipole kicks, we investigated this growth rate to see if we can place an upper-bound on the level of dipole kicks and consequently the path-length jitter in the bypass caused by them.

In \cite{Val_Noise} the emittance growth from turn-by-turn uncorrelated dipole errors was computed. In CESR the revolution frequency is 390.1 kHz so that turn-by-turn uncorrelated errors are not realistic. However, following their same approach we estimate the emittance growth using, instead, a hard-change model where a dipole error is kept constant for $N$ turns and then abruptly changes to a new random uncorrelated value. In this case the correlation function between turns $n$ and $m$ at the $k^{th}$dipole is $\langle R_{err,n}\theta_{k},R_{err,m}\theta_{k}\rangle=R_{err,n}^2\theta_{k}^2$ for $n,m<N$, and $\langle R_{err,n}\theta_k,R_{err,m}\theta_k\rangle=0$ for $n,m\geq N$. We find that, each time a dipole is changed, after decoherence and neglecting synchrotron damping, the emittance will grow by
\begin{equation}
 \Delta \epsilon_k=\beta_k\frac{R_{err,k}^2\theta_k^2}{4\sin^2(\pi \nu)}\bigg(\cos^2(\pi\nu)+1/2\bigg).
\end{equation}
Then the average growth rate is found by summing over all dipoles and dividing by the period between changes
\begin{equation}
    \frac{d\epsilon}{dt}\approx\frac{\sum_k\epsilon_k}{NT}
\end{equation}
where $T$ is the revolution period. 

\begin{figure}
	\centering
	\includegraphics*[width=0.485\textwidth]{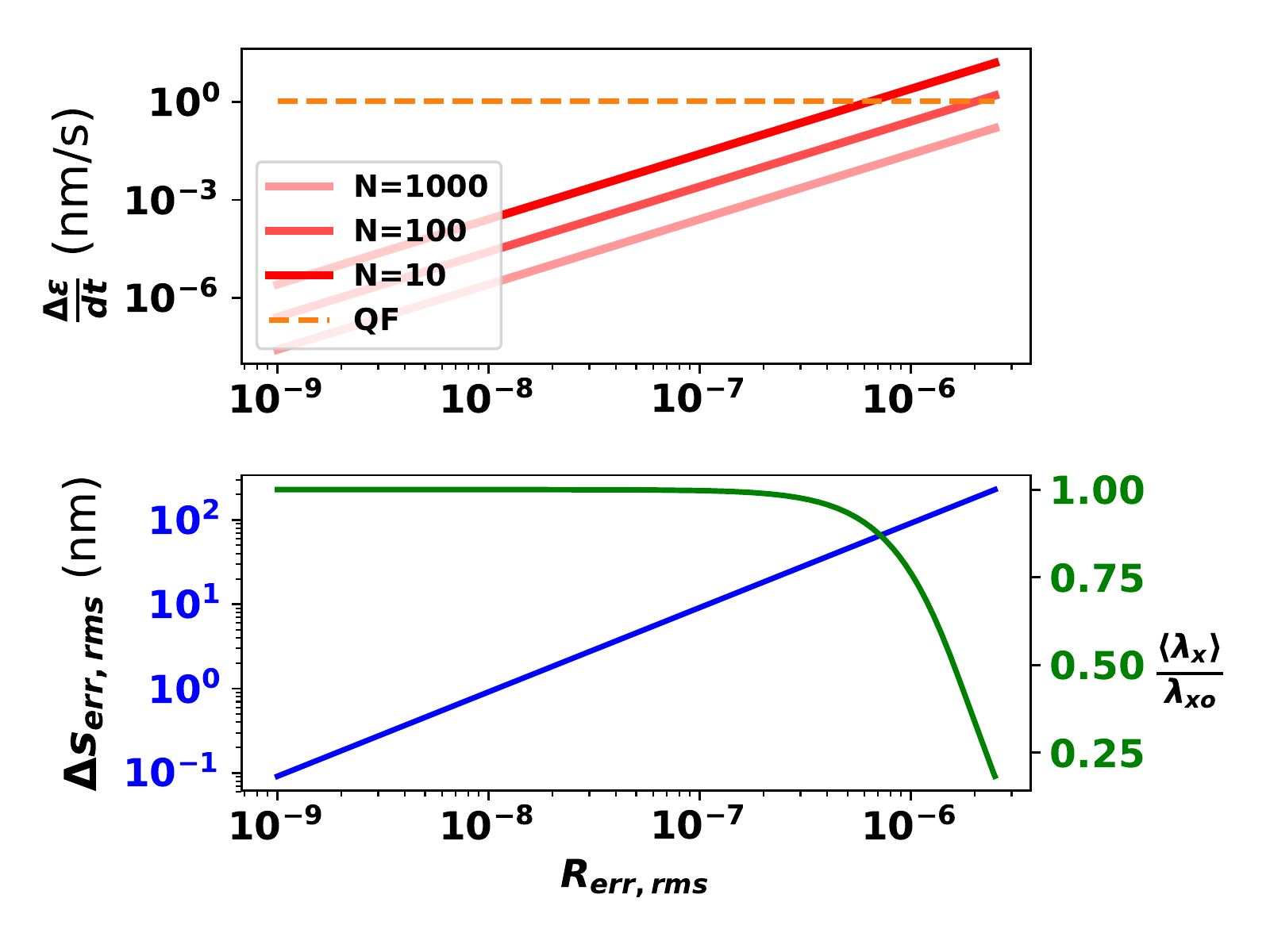}
	\caption{Top: Computed emittance growth rates from dipole fluctuations as a function of ring wide RMS relative dipole error and the rate from QF of synchrotron radiation. Bottom: The expected RMS path-length jitter (blue) and OSC damping rate reduction (green) as a function of RMS relative dipole error. }
	\label{Fig:EmittanceGrowth}
\end{figure}

The top plot of Fig.~\ref{Fig:EmittanceGrowth} shows the emittance growth for $N=10,100,1000$ as a function of the RMS relative error of the dipoles, $R_{err,rms}$. For comparison the growth rate from synchrotron radiation Quantum Fluctuations (QF) is also plotted. The growth rate is inversely dependent on $N$ and even in the extreme case of $N=10$, in order that growth from dipole noise remains much smaller than growth from QF (say less than 1$\%$) $R_{err,rms}<2.0\times 10^{-7}$. At that noise level the reduction in OSC damping rates from dipole induced path-length jitter, does not exceed 2 $\%$ which can be seen from the bottom plot of Fig.~\ref{Fig:EmittanceGrowth}.

The above analysis would seem to indicate dipole noise levels are too small to substantially affect the OSC process. However, dipole noise can come from power supply ripple which is a continuous waveform comprised of discrete frequencies---for example 60 Hz and its first few harmonics. In a case like this, it is also shown in \cite{Val_Noise} that noise produces emittance growth only when its spectral density overlaps with a side-band of the betatron frequency
\begin{equation}
    \frac{d\epsilon}{dt}=\frac{1}{4\pi}\sum(\nu)\Omega^2
\end{equation}
where $\Omega=2\pi/T$ is the angular revolution frequency and
\begin{equation}
    \sum (\nu)=\beta_k\bigg(\frac{qL_k}{U_s}\bigg)^2\sum_{n=-\infty}^{n=\infty}S_{\delta B}((\nu-n)\Omega)
\end{equation}
where $n$ is an integer, $\nu$ is the tune and $S_{\delta B}(\omega)$ is the spectral density of the dipole field noise. In CESR power supply ripple is likely more than two orders of magnitude less than the revolution frequency. Consequently, in order for a betatron side-band to overlap with it, the fractional part of the tune would need to be within less than 1 $\%$ of an integer. Obviously, such a tune is avoided in a storage ring and therefore, noise from power supply ripple does not produce emittance growth. It can however, still produce degrading path-length jitter if it causes the dipole current to fluctuate $\gtrapprox 5\times 10^{-7}$.

Furthermore, noise can come from stray magnetic fields of other electrical components near the bypass and may produce sizeable path-length jitter. For example, during a CESR machine study a technique to identify transverse kicks to the beam orbit~\cite{BerganSpeedofLight} found an approximately 1.3 $\mu$rad horizontal kick with a 180 Hz frequency near the center of the bypass. This kick would have resulted in an approximately 5 $\mu$m path-length fluctuation---essentially reducing the damping rate to zero. Fortunately, the source of the stray field was identified as a problematic power supply which has since been fixed. These considerations lead us to conclude we can not rule out dipole noise as a source of path-length jitter based on the machine emittance. 
\subsection{Quadrupole Motion}
A transverse displacement $\Delta d$ of a quadrupole produces a constant magnetic field error, $B_{err}=\Delta d B'$, where $B'$ is the gradient of the quadrupole, which is related to the focal length as $f=cP/eB'l_{quad}$ with $l_{quad}$ being the length of the quadrupole. Using a similar approach from Section~\ref{Sec:StabilityIncoherent} and for brevity now only considering quadrupoles inside the bypass, the path-length error is found to be
\begin{equation}
\Delta s_{quad}=\frac{\Delta d}{f}(M_{51,quad}l_{quad}+M_{52,quad})    
\end{equation}
where, as before, the transfer elements are evaluated from the exit of the quadrupole to the KU center. Again because $M_{51,quad}$ and $M_{52,quad}$ grow quite large in the bypass, the path-length is very sensitive to quadrupole vibrations. For example, if each quadrupole in the bypass vibrates independently with an RMS value of 50 nm, the expected RMS path-length jitter will be 100 nm. 

When considering dipoles, we also needed to account the additional path-length accrued by the particle as it travelled through the element using Eq.~\ref{s1_incoh}. For a quadrupole this term is
\begin{equation}
\Delta s_{quad,2}=\frac{l}{6}\frac{\Delta d^2}{f^2}.
\end{equation}
The quadratic dependence in $\Delta d/f$ arises because both $\Delta x$ and $\rho$ are generated from $\Delta d$. Since $\Delta d \ll f$ this term is negligibly small. 

Finally, the use of mirrors in the light transport will add an additional source of path-jitter through mechanical vibrations that can be $\mu$m scale in magnitude. Accounting these noise sources, a feedback system for path-length stabilization will be a critical component for our experiment and is investigated in the next section.
\section{Feedback for path length stability}
\label{Sec:Feedback}
OSC works by appropriately modulating the relative path-length of each particle from PU to KU to provide a corrective energy kick. In the absence of path-length errors the reference particle arrives at the zero crossing of its PU wave-packet and receives no kick, while all other particles in the bunch are delayed or advanced, with respect to their own PU-wavepackets, and either constructively or destructively interfere with their KU wave-packets. The bunch centroid coincides with the reference particle so that there are an equal number of particles adding constructively and destructively; the total energy radiated in the PU and KU by the bunch will appear to be the sum of both undulators in the \textit{absence} of interference---a path-length error breaks this symmetry. Consequently, there will be either an enhancement or reduction of the far-field radiation that allows the arrival phase of the bunch centroid to be measured \cite{Andorf_NAPAC} to the precision required to obtain near ideal (error-free) damping. The total radiated energy $\Delta E_{beam}$ (in the first harmonic of the undulators) as the path-length is varied, is found by averaging the combined radiated energy of each particle in the PU and KU. For a Gaussian beam
\begin{equation}
\begin{aligned}
\Delta E_{beam}=-N_p\Delta \mathcal{E}\bigg[1+
\frac{s_{pulse}-\vert s_{err} \vert}{s_{pulse}}\sin(k_ls_{err})&\\
\times\exp\bigg(-\frac{\mu_{0,1}^2}{2}\bigg(\frac{1}{\eta_x^2}+\frac{1}{\eta_p^2}\bigg)\bigg)\bigg]
\end{aligned}
\end{equation}
where $s_{pulse}\approx N_u\lambda_l$ is length of the undulator wave-packet and $N_p$ is the number of particles per bunch. The visibility of the energy modulation is determined by the exponential term, which in the CESR demonstration will be $\approx 0.72$. 

In order to accurately measure the phase, the average power of the interfered PU and KU signals must be larger than the detector's minimum detectable power at maximum destructive interference (i.e. $k_ls_{err}=-\pi/2$). We envision performing this measurement with a fast photo-diode, and for a concrete example, consider a silicon-based photo-diode (model DET025A from Thorlabs). With a response time of 150 ps and a noise equivalent power given in the data sheet\cite{Thorlabs}, the minimum detectable power is found to be 0.5 nW. The radiation from the PU and KU is emitted over a bandwidth comparable to that of the diode's spectral response. To account this, Synchrotron Radiation Workshop (\texttt{SRW})\cite{SRW} was used to compute the radiation spectrum of a single undulator. The product of this spectrum and a normalized spectral response of the diode was integrated to get an effective energy measurement. The finite spectral response is expected to increase the minimum detectable power by $\approx 0.54$ and we additionally included another $50\%$ increase as an estimate of the quantum efficiency. We find that for an $S/N=10$ at the minimum detectable power, we require $N_p=2.7\times 10^6$ particles per bunch.
\begin{figure}
	\centering
	\includegraphics*[width=0.485\textwidth]{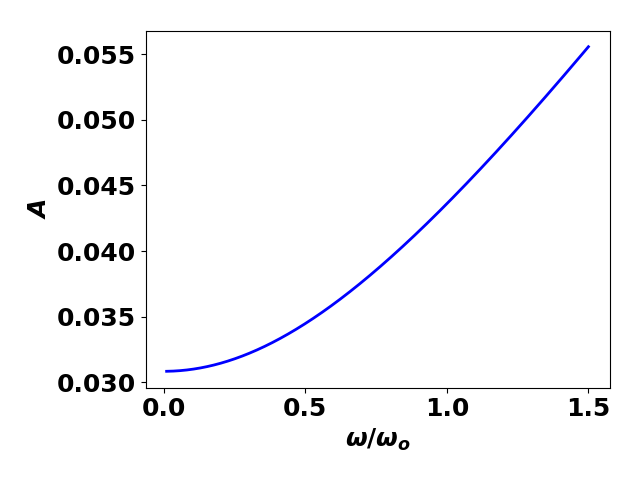}
	\caption{Reduction in the path-length error amplitude as a function of noise frequency for an EOM-based feedback system.}
	\label{FeedbackFig}
\end{figure}
Because the arrival time of the signal is well known, signal gating can be used to further decrease the required number of particles. 50\,ns gating is a reasonably large window and decreases the integrated noise spectrum by a factor $\approx 50$. Therefore, the minimum required number of particles for a phase-measurement is $N\approx 5\times 10^4$. This number corresponds to a beam current much smaller than the 1 $\mu$A (set to limit emittance growth from IBS) anticipated for the OSC demonstration.

The above measurement can provide turn-by-turn information of the arrival phase of the bunch centroid that can be used for feedback. In principle anything that alters either the light or particle path transit-time (e.g. a movable mirror or additional dipole corrector in the bypass) can be used for feedback. Here we consider the use of an Electo-Optic Modulator (EOM). An EOM is an electric-optic device that induces a change in the index of refraction of a nonlinear crystal by applying a voltage, and consequently modifies the time-of-flight of light passing through it. EOMs can operate over a large spectral range as they are often used with ultra-fast lasers, can be modulated from DC to several 100 kHz, and can provide a few microns of path-length modulation. 

We specifically consider an EOM with 10 kHz bandwidth and a path-length adjustment range of $5\lambda$. In Appendix \ref{Sec:Appendix} expressions (Eq.'s~\ref{SinSolution} and \ref{CosSolution}) for the feedback corrected path error $\Delta s_{err,cor}(t)$, are given for initial path-length errors $\Delta s_{err}(t)=\Delta s_o\sin(\omega t)$ and $\Delta s_{err}(t)=\Delta s_o\cos(\omega t)$, that consist of a fast decaying transient component and an oscillating component of amplitude $A\Delta s_o$ where $A$ is given in Eq.~\ref{A}. Typically the transient term's decay time is much less than one period of oscillation and so can be ignored. In this case $A$ is simply the ratio of corrected to uncorrected path length error and is plotted in Fig.~\ref{FeedbackFig} for our assumed EOM parameters. Preliminary work has shown 60 Hz line ripple and its first few harmonics (particularly the 3rd) are the strongest source of dipole noise in CESR and thus from Fig.~\ref{FeedbackFig} an EOM based feedback system reduces sensitivity to this noise by approximately a factor of 30. 
\section{Multi-Particle Tracking Simulation of OSC}
\begin{figure}
	\centering
	\includegraphics*[width=0.485\textwidth]{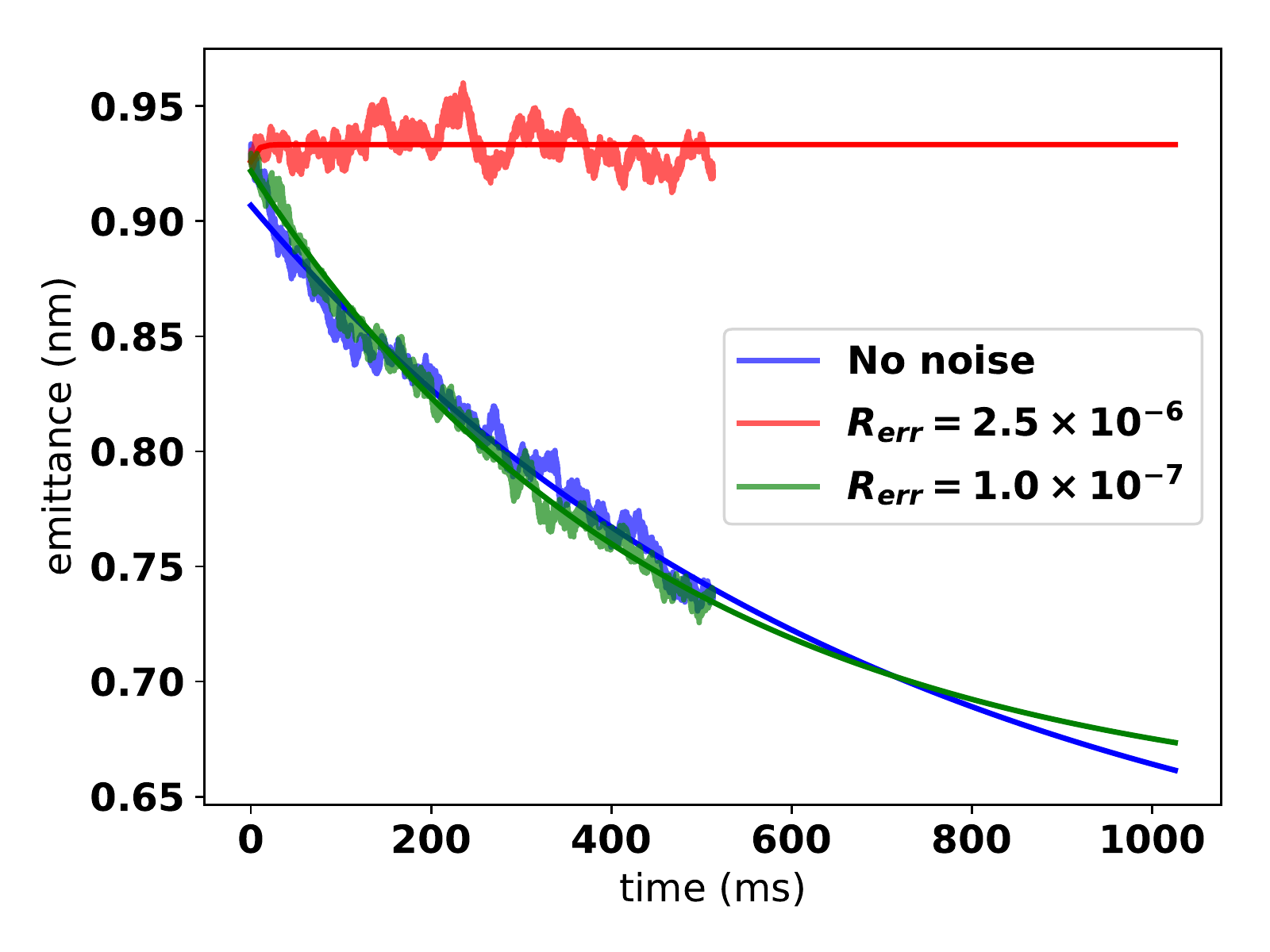}
	\caption{Particle tracking results for OSC \texttt{BMAD} simulations that include both OSC and SR damping, QF and OSC incoherent kicks. Various levels of dipole noise are included. The solid lines are exponential best fits.}
	\label{Fig:DampingFig}
\end{figure}
As a final analysis of the arc-bypass, and to confirm the predicted dipole stability requirement found above, we present particle tracking simulations of the OSC process that include both the coherent and incoherent contributions of the kick imparted to a particle, as well as synchrotron radiation effects (damping and excitation). Implemented with \texttt{BMAD}\cite{BMAD} routines, the simulation tracks 1000 macro-particles through the CESR lattice, with a distribution corresponding to the emittance set by synchrotron radiation in the absence of OSC. 

OSC is included by recording the transit-time from PU to KU of each particle, relative to the reference orbit transit-time. The particle's energy is changed according to Eq.~\ref{deltau}. In reality the particle receives a kick over the length of the KU, but in simulation, as a simplification, we apply the kick as an impulse at the center of the KU. The applied kick so far represents the coherent cooling process; we additionally now include an estimate of the incoherent heating effect arising from kicks applied by neighboring particles within a longitudinal slice of the beam, of a width typically estimated\footnote{This estimate neglects that only a few cycles of the wave-packet are in focus at a given longitudinal point in the KU and, therefore, can overestimate the longitudinal slice length by a non-negligible amount, see \cite{Andorf_NIM} for details. For this reason, when considering incoherent kicks, we assume a 4 period undulator.} as $N_u\lambda_o$. The effect of the incoherent kicks depends on the number of particles within a slice. To directly compute this effect, the arrival time of all $10^7$ electrons anticipated for our demonstration would need to be tracked, which is not practical. A numerical investigation showed that, as long as the number of particles per slice, $N_s\gtrapprox$ 6, the incoherent kicks received by any particular particle is approximately Gaussian random, with an amplitude and width, $N_s(\frac{\Delta\mathcal{E}}{U_s})^2 $ and $\sqrt{N_s}$ respectively. Incorporating the incoherent kicks in this way allows for a fast and accurate estimate of the heating term that does not depend on the number of macro-particles used in the tracking.

Although the major thrust of the OSC program in CESR is the demonstration of active OSC, observing passive cooling will be an important program milestone, and so in these simulations we use a kick amplitude corresponding to the passive value of 420 meV. As a baseline we first performed a simulation without bend noise. The tracking result is shown in the blue trace of Fig.~\ref{Fig:DampingFig} with damping clearly visible. An exponential fit was performed and indicates that OSC will reduce the equilibrium emittance by 30$\%$.

Next we include bend noise as random Gaussian fluctuations in field strength. The hard-change model was used, with the dipole errors updated every 100 turns. We choose to update every 100 turns since, based on the analysis in \ref{Sec:Stability}, a sufficiently large RMS relative error ($2.5\times 10^{-6}$) could be applied so that the path-length jitter would significantly reduce the damping effect, without also introducing a sizable emittance growth. To further suppress the emittance growth, noise was applied only to dipoles inside the bypass.

Using the formulas developed in Section~\ref{Sec:Stability}, we predict that a relative error of $10^{-7}$ will reduce the damping rate by 1 $\%$, while an error of $2.5\times10^{-6}$ reduces it by 80 $\%$. The results are shown in Fig.~\ref{Fig:DampingFig}. As expected for an error of $10^{-7}$, the green trace shows only a marginal difference in the damping rate and equilibrium emittance, while for an error of $2.5\times10^{-6}$, shown in red, OSC damping is no longer visible.
\label{Sec:Sims}
\section{Conclusion}
\label{Sec:Conclusion}
We have presented an arc-bypass concept for a demonstration of OSC in CESR that could be an effective technique for cooling high energy hadron beams. The major advantage of the arc-bypass, as compared to the dog-leg chicane bypass, is the large optical delay of approximately 20 cm (in CESR's case) which enables either multi-pass or staged amplification schemes for the OA. The many centimeter relative delay is essential in an implementation where high gain amplification is required, as is the case for hadron and heavy-ion cooling scenarios. We have characterized the stability requirement of dipoles in the bypass and provide formulas to compute the path-length error in terms of dipole field errors,  and also an estimate on the reduction of the OSC damping rates resulting from path-length error. We find that the arc-bypass is sensitive to dipole fluctuations since (i) it uses stronger bending magnets than are typically required for a dog-leg chicane and (ii) the long separation between the PU and KU allows for $M_{51}$ and $M_{52}$ to grow quite large in the bypass. We investigated a feedback system to compensate changes in the path-length from the PU to KU in order to relax tolerances on dipole stability. Finally, particle tracking simulations were performed to demonstrate the dependence of OSC damping rates on various levels of bend noise, in the absence of path-length correction.
\section{Acknowledgements}
This work was supported by the U.S. National Science
Foundation under Award No. PHY-1549132, the Center
for Bright Beams, NSF-1734189,  DGE-1144153.
\appendix
\section{Analytic Feedback Solution}
\label{Sec:Appendix}
In this appendix expressions for the feedback corrected path-length error assuming an initially uncorrected oscillating path error are derived. As was presented in section \ref{Sec:Feedback}, a path-length error $\Delta z_{err}$\footnote{For this appendix we have changed our symbol for path-length error from $\Delta s$ to $\Delta z$ to reserve $s$ for its standard use as the Laplace domain variable.} can be inferred by measuring the total energy radiated from the PU and KU in the first harmonic of the radiation. Neglecting the finite length of the undulator wave-packet a photo-diode registering the total energy will have a voltage readout
\begin{equation}
V_{out}=K_1\sin(k_l\Delta z_{err})
\label{Voutsin}
\end{equation}
where $K_1$ is an arbitrary proportionality constant that converts energy to voltage. In control theory the element or process that converts the system input ($\Delta z_{err}$) to the output signal ($V_{out}$) is called the \textit{plant} and denoted as $P$. Likewise, the element or elements that act on the plant's output in order to make a feedback correction is called the \textit{controller} and denoted as $C$. For a linear plant and controller the feedback corrected output is related to the input as
\begin{equation}
V_{out}(s)=\Delta z_{err}\frac{P(s)}{1+P(s)C(s)}
\end{equation}
where $s=\sigma +i\omega$ is a complex frequency in the Laplace domain. In general our plant is non-linear; However, in order to apply some elementary control theory we can linearize the plant by assuming $\sin(k_l\Delta z_{err})\approx k_l\Delta z_{err}$  so that
\begin{equation}
P(s)=K_1
\end{equation}
and for tidiness we absorbed $k_l$ into $K_1$.

\begin{figure}
	\centering
	\includegraphics*[width=0.475\textwidth]{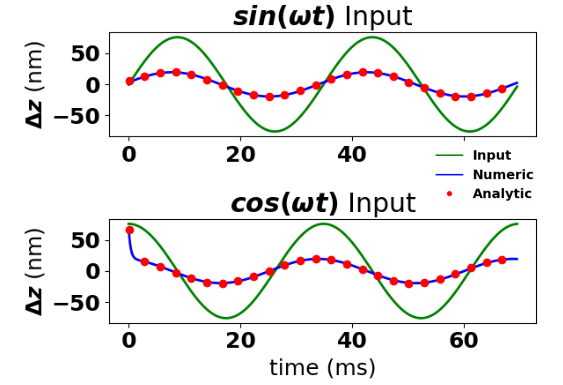}
	\caption{Example solutions of the feedback corrected path-length error.}
	\label{FeedbackSolFig}
\end{figure}

We consider a proportional controller with its output passed through a low-pass filter in order to model the finite speed that the path-length can be corrected
\begin{equation}
C(s)=\frac{k_p}{1+s/\omega_o}
\end{equation} 
where $k_p$ is an adjustable proportionality constant of the controller and $\omega_o$ is the cut-off frequency. For an input $\Delta z_{err}=z_o\sin(\omega t)$, $V_{out}$ becomes
\begin{equation}
V_{out}(s)=\Delta z_o\frac{\omega}{s^2+\omega^2}\times\frac{K_1}{1+\frac{K_1k_p}{1+s/\omega_o}}.
\end{equation}
After converting the above expression into the time domain, the feedback corrected error can be written as
\begin{equation}
\Delta z_{err,cor}(t)=\Delta z_o\big(A\sin(\omega t+\delta_{sin})+B\exp{(-\alpha t)}\big)
\label{SinSolution}
\end{equation}
where
\begin{equation}
A=\sqrt{\frac{1+(\omega/\omega_o)^2}{1+(\omega/\omega_o)^2(1+K_1k_p)^2}},
\label{A}
\end{equation}
\begin{equation}
B=\frac{\omega}{\omega_o}\frac{K_1k_p}{(\omega/\omega_o)^2+(1+K_1k_p)^2},
\end{equation}
\begin{equation}
\delta_{sin}=\tan^{-1}(\chi)=\tan^{-1}\bigg(\frac{\omega}{\omega_o}\frac{K_1k_p}{(\omega/\omega_o)^2+(1+K_1k_p)}\bigg)
\end{equation}
and 
\begin{equation}
\alpha=\omega_o(1+K_1k_p).
\end{equation}
For an input $\Delta z_{err}=z_o\cos(\omega_t)$, we find a similar solution
\begin{equation}
\Delta z_{err,cor}(t)=\Delta z_o\big(A\cos(\omega t+\delta_{cos})+C\exp{(-\alpha t})\big)
\label{CosSolution}
\end{equation}
with
\begin{equation}
C=\frac{K_1k_p(1+K_1k_p)}{(\omega/\omega_o)^2+(1+K_1k_p)^2}
\end{equation}
and
\begin{equation}
\delta_{cos}=-\tan^{-1}\bigg(\frac{1}{\chi}\bigg).
\end{equation}
Example plots of Eq.'s~\ref{SinSolution} and \ref{CosSolution} are shown in red in Fig.~\ref{FeedbackSolFig} for parameters, $z_o=75$ nm, $\omega=180$,Hz $\omega_o=630$ Hz, $K_1=1.0$ and $k_p=3.0$. A numeric solution was obtained using the \textit{Python Controls System Library} shown in blue and is seen to be in good agreement with the analytic solutions.

In their current form these two solutions contain two arbitrary constants $K_1$ and $k_p$. For a linear model there is no bound on $V_{out}$ or $k_p$. In reality, the sinusoidal nature of the plant implies $V_{out,max}=K_1$. Additionally, the controller will have a finite range over which it can make a path correction, $\Delta z_{feed,max}=N\lambda_l$ where $N$ is the number of wavelengths of the range. Thus, the maximum value of $k_p$ is such that when $V_{out}=K_1$,~$\Delta z_{feed}=N\lambda_l$. At this maximal the product $K_1k_p=2\pi N$ and the above expressions for the feedback corrected path are seen to depend on three values only, $\omega$, $\omega_o$ and $N$.

None of the expressions above are valid if $k_l\Delta z_{err}$ is too large and the linearity assumption of the plant becomes invalid. Therefore, we first require a small initial path-length error; then the condition for the error to remain small is $\frac{\Delta z_{err}}{dt}\bigg\vert_{max}<\frac{\Delta z_{feed}}{dt}\bigg\vert_{max}$. Thus, the linearity of the plant is valid if either
\begin{equation}
\Delta z_{o}<\frac{\omega_o}{\omega}N\lambda_l, \quad \omega>\omega_o
\end{equation}
or
\begin{equation}
\Delta z_{o}<N\lambda_l, \quad \omega<\omega_o.
\end{equation}


\begin{thebibliography}{apssamp}
\bibitem{Mohl}
D. M{\"o}hl, G. Petrucci, L. Thorndale, S. van der Meer \href{https://doi.org/10.1016/0370-1573(80)90140-4}{Physics Reports \textbf{58} (2) p.73-119 (1980)}.
\bibitem{OSC_Mikhailichenko}
	A. A. Mikhailichkenko, M. S. Zolotorev, \href{https://doi.org/10.1103/PhysRevLett.71.4146}{Phys. Rev. Lett. {\bf 71} (25),  4146 (1993)}.
\bibitem{OSC_Zolotorev}
M. S. Zolotorev, A. A. Zholents, \href{https://doi.org/10.1103/PhysRevE.50.3087}{Phys. Rev. E, {\bf 50} (4),  3087 (1994)}.
\bibitem{SYLee}
S.Y. Lee, Y. Zhang, K.Y. Ng \href{https://doi.org/10.1016/j.nima.2004.06.063}{Nucl. Instrum. Meth A {\bf 532},(2004)}.
\bibitem{Pasquinelli}
R. J. Pasquinelli, JINST 6 T08002 (2011)
\bibitem{CEC1}
V.N. Litvinenko, Y.S. Derbenev, \href{https://doi.org/10.1103/PhysRevLett.102.114801}{Phys. Rev. Lett., {\bf 102} (114801) (2009)}.
\bibitem{MBEC1}
D. Ratner, \href{https://doi.org/10.1103/PhysRevLett.102.114801}{Phys. Rev. Lett., {\bf 102} (084802) (2013)}.
\bibitem{IOTAVal} V. Lebedev, Yu. Tokpanov and M. Zolotorev, in Proc. of the North American Particle Accelerator Conference (NAPAC'13), Pasadena, CA, U.S.A., 29 Sept.-4 Oct. 2013, p. 422 (2013).
\bibitem{Jarvis} J.D. Jarvis et al \href{https://doi.org/10.18429/JACoW-HB2018-TUP2WA01} {in Proc. of the ICFA Advanced Beam Dynamics  Workshop on High-Intensity and High-Brightness Hadron Beams (HB2018), Daejeon, Korea , 17-22 June 2018, p. 168-173 (2018).}
\bibitem{OSC_val} V.A. Lebedev, ``Optical Stochastic Cooling" in ICFA Beam Dyn.Newslett. {\bf 65} pp. 100-116 (2014).
\bibitem{Amplifier_Andorf}
M.B. Andorf et al, \href{https://doi.org/10.1364/CLEO_AT.2017.JW2A.90}{ in Conference on Lasers and Electro-Optics, OSA Technical Digest (online) (Optical Society of America, 2017), paper JW2A.90.}
\bibitem{Amplifier_Zholents}
A. A. Zholents, M. S. Zolotorev, \href{https://doi.org/10.1109/PAC.1997.751023}{in Proc. of the Particle Accelerator Conference (PAC'97), Vancouver, B.C, Canada, p. 1804 (1997).}
\bibitem{Andorf_NIM} M.B. Andorf, V. A. Lebedev, P. Piot, J. Ruan,  \href{https://doi.org/10.1016/j.nima.2017.11.094}{Nucl. Instrum. Meth A {\bf 883}, 119 (2018)}. 
\bibitem{Andorf_PRAB}
	M.B. Andorf et al, \href{https://doi.org/10.1103/PhysRevAccelBeams.21.100702}{Phys. Rev. Accel. Beams, {\bf 21}, 100702 (2018)}.
\bibitem{Andorf:Thesis} 
  M.~B.~Andorf,
  \href{https://doi.org/10.2172/1462087}{Ph.D. thesis, Northern Illinois University, 2018, ``Light Transport and Amplification for a Proof-of-Principle Experiment of Optical Stochastic Cooling''.}
 \bibitem{Tao}
D. Sagan, J. Smith, in the Proc. of the Particle Accelerator Conference (PAC'05), Knoxville, TN, USA, 16-20 May 2005. pp. 4159-4161.
\bibitem{Bergan_IPAC}
W.F. Bergan, et al, \href{https://doi.org/10.18429/JACoW-IPAC2019-MOPGW100}{in the Proc. of the International Particle Accelerator Conference (IPAC'19), Melbourne, Australia, 19-24 May 2019, p. 360-363.}
\bibitem{VAL_COOL15}
V.A. Lebedev, A.L. Romanov, in Proc. of the International Workshop on Beam Cooling and Related Topics (COOL'15), Richmond VA, USA, p. 123 (2015)
\bibitem{Kafka:Thesis}
G. Kafka, Ph.D. thesis, 2015, Illinois Institute of Technology, ``Lattice Design of the Integrable Optics Test Accelerator and Optical Stochastic Cooling  Experiment at Fermilab''.
\bibitem{Bengstton}
J. Bengsston, ``The sextupole scheme for the Swiss Light
Source (SLS): an analytic approach”, PSI, Villigen, Switzer-
land, SLS Note 9/97, 1997.
\bibitem{ZholentsErrors}
A. A. Zholents, M. S. Zolotorev, \href{https://doi.org/10.1109/PAC.1997.751022}{in Proc. of the Particle Accelerator Conference (PAC'97), Vancouver, B.C, Canada, p. 1801 (1997).}
\bibitem{Wiedemann}
H.Wiedemann, \textit{Particle Accelerator Physics}, 3rd ed. Springer (2007). 
\bibitem{Meller}
R.E. Meller et al, Report No. SSC-N-360, 1987. 
\bibitem{Ng}
K.-Y. Ng, J.M. Peterson, \href{ doi: 10.1109/PAC.1991.164731}{in Proc. of the IEEE Particle Accelerator Conference, San Francisco, CA, USA, pp. 1645-1647 (1991).} 
\bibitem{Val_Noise}
V. Lebedev et al, \textit{Particle Accelerators} {\bf44} p. 147-164, (1994).
\bibitem{BerganSpeedofLight}
W. Bergan et al, \href{https://doi.org/10.1103/PhysRevD.101.032004}{Phys. Rev. D., {\bf 101} 032004 (2020)}
\bibitem{Andorf_NAPAC}
M.B. Andorf et al, \href{https://doi.org/10.18429/JACoW-NAPAC2016-WEPOA38} {in Proc. of the North American Particle Accelerator Conference (NAPAC'16), Chicago, IL, U.S.A., October 9-14. 2016, p. 779-781}.
\bibitem{Thorlabs}
Thorlabs, \href{https://www.thorlabs.com/drawings/2b77b3ebd00c3fe5-E71E4B5E-CD27-56EA-3B7F9D8427C64F3A/DET025A-Manual.pdf}{Det025A Free Space Window Input Si Biased Detector User Guide} (2017).
\bibitem{SRW}
O. Chubar, P. Elleaume, in the Proceedings of the European Particle Accelerator Con-
ference (EPAC’98), Stocholm, Sweden, p.1177 (1998).
\bibitem{BMAD}
D. Sagan, \href{https://doi.org/10.1016/j.nima.2005.11.001}{in the Proceedings of the International Computational Accelerator Physics Conference (ICAP'04), St.Petersburg, Russia, June 29-July 2 (2004).}
\end{thebibliography}
\end{document}